\documentstyle[times,namedreferences]{kluwer}
\begin{opening}
\title{Attributes of Gravitational Lensing Parallax}
\author{Robert J. \surname{Nemiroff}}
\institute{Department of Physics, 
Michigan Technological University,
Houghton, MI 49931, USA}
\date{}
\end{opening}

\begin{document}

\begin{abstract}
The density of stars and MACHOs in the universe
could theoretically be determined or limited by simultaneous 
measurements of compact sources by well separated observers.  
A gravitational lens effect would be expected to create a slight 
differential amplification between the observers detectable with 
sufficiently sensitive relative photometry: ``lensing parallax."   
When applied to expanding fireballs such as those from GRBs 
and supernovae, the mass of the lens can be indicated by the 
end of lensing parallax, when the angular size of the source 
becomes much greater than the angular size of the Einstein ring 
of the lens. 
\end{abstract}

\section{Introduction}

The cosmological density of compact objects in 
the range $10^{-15}$ - $10^{10}$ is currently uncertain.  
Potentially important components to the density of the universe  
include stars, MAssive Compact Halo Objects (MACHOs), 
and massive black holes.  Here one 
hypothetical path to estimating the space density of these objects  
is proposed: the simultaneous measurement by separated 
observers of compact sources at cosmological distances.  A 
population of compact objects of
$\Omega_{L}$ would cause a slight gravitational lens 
induced brightness difference that would be recoverable by 
sufficiently sensitive photometry.  As this method hinges on 
well calibrated observations by separated observers, it will 
be referred to as ``gravitational lensing parallax" or just 
``lensing parallax."
 
Lensing parallax observations of cosmological sources were first 
suggested by Refsdal (1966), who discussed what generic 
information about lensing events could be obtained by two 
platforms separated in the solar system.  That lensing parallax 
observations of QSOs could determine the transverse velocity
of galaxies was pointed out by Grieger, Kayser \& Refsdal 
(1986).  Gould (1992, 1994, 1995) suggested the method could better 
constrain the transverse velocities of lenses in MACHO 
searches.  

GRBs were discovered in the late 1960s (Klebesadel, Strong, \& 
Olson, 1973).  In 1997, GRBs were found to have
counterparts in wavelength bands from the radio to the X-ray 
(Costa et al., 1997; van Paradijs et al., 1997).  For simplicity, 
these event will all be referred to as GRBs, however.  In 
general, GRB emission lasts longer and peaks later at longer 
wavelengths.  At the time of this writing, one GRB was 
found to have an absorption line 
placing the event at cosmological distances (Metzger et al. 
1997).  

Microlensing of GRBs had been proposed by Paczynski (1986).
A search for gravitational lensing effects in GRB data has been
ongoing since 1993 (Nemiroff et al., 1993; 
Marani, 1997; Marani et al., 1998).  
Nemiroff \& Gould (1995) first suggested lensing parallax 
observations of gamma-ray bursts (GRBs), and showed how 
limits could be placed on a cosmological abundance of lens 
masses from $10^{-15}$ $M_{\odot}$ to $10^{-7}$ 
$M_{\odot}$. Loeb \& Perna (1998) recently
discussed how lensing parallax observations of GRB afterglows 
could confirm a microlensing amplification bump or a lens-
induced polarization signature.
 
In this paper the initial lensing parallax analysis
of Refsdal (1966) is expanded to focus on its ability to detect 
a potential
cosmological abundance of compact dark matter candidates.  
In particular, the paper explores whether lensing parallax might 
lead to a useful probe of a newly discovered type of sources:
GRB afterglows, by a newly reported type of lenses: MACHOs
(Alcock et al., 1996).

\section{The Theory and Practice of Gravitational Lensing 
Parallax}
 
The amplification of a point source by a compact lens is
 \begin{equation}
 A = {u^2 + 2 \over
       u \sqrt{u^2 + 4}},
 \end{equation}
where $u$ is the angular separation between the lens and source 
on the sky of the observer, in units of $\theta_E$, the angular 
Einstein ring size of the lens. It is also well known that 
 \begin{equation}
 \theta_E = \sqrt{2 R_S D_{LS}\over
                          D_{OL}D_{OS}},
 \end{equation}
where $R_S$ is the Schwarzschild radius of the lens, $D$ is 
angular diameter distance, and $O$, $L$, and $S$ subscripts 
designate observer, lens, and source, respectively.  However, 
equation 1 also holds when $u$ is taken to be projected distance 
in the observers plane, measured in terms of $E_O$ the Einstein 
ring projected into the observer's plane (Nemiroff \& Gould, 
1995), where
 \begin{equation}
 E_O = \sqrt{2 R_S D_{OL}D_{OS}\over
                         D_{LS}}.
 \end{equation}

Following Refsdal (1966), differential amplification is investigated.  
From equation 1 it is straightforward to show that 
 \begin{equation}
| {dA \over du} | = {8 \over u^2 (u^2 + 4)^{3/2}}.
 \end{equation}
 
Suppose two observers are separated in the solar system by an 
amount $\Delta u$.  They will detect a lensing parallax event 
when the difference in the apparent luminosity each measures
is greater than $\Delta A$.   Here two detection scenarios will
be considered: small $u$ events, and large $u$ events. 
 
Small $u$ events corresponds to high $A$.  Equation 4 can then be 
approximated as 
 \begin{equation}
  \Delta A \sim {dA \over du }\Delta u \sim 
  {\Delta u \over u^2}.
 \end{equation}
Therefore
 \begin{equation}
 u^2 \sim {\Delta u \over \Delta A }
 \end{equation}
when $u < 1$.  

Now the probability of lensing above any amplitude is given by 
Nemiroff (1989) for a smooth $\Omega=1$ universe as
 \begin{equation}
  P = 6 \Omega_L \Phi \Psi
 \end{equation}
where $\Phi = \sqrt{A^2/(A^2 - 1)} - 1$, and 
\begin{equation}
 \Psi = \left( { 4 + S^{1/2} \ {\rm ln} \ S -
 4 S^{1/2} + \ {\rm ln} \ S \over
    S^{1/2} - 1 } \right) ,
 \end{equation}
where $S = (1 + z_S)$.  Nemiroff (1989) also showed that 
$u^2 = 2 \Phi$, so that $P = 3 u^2 \Omega_L \Psi$.  
Using equation 6 it is found that 
 \begin{equation}
 P \sim 3 \ \Omega_L \Psi 
       \left( { \Delta u \over \Delta A } \right) .
 \end{equation}

To translate this relation to a ratio of more intuitive values, it will
be assumed that the source is at redshift $z_S = 1$, so that $\Psi \sim 0.04$ 
and, taking $H_o = 65$ km sec$^{-1}$ Mpc$^{-1}$, $D_{OS} = (2c/H_o) (S^{-1} - S^{-3/2}) = 1.35$ Gpc.  It will also be assumed here that the lens is placed 
at $D_{OL} / D_{LS} = 1$, near its most probable position, and
that $\Delta u = D_{sep}/E_O$.  From $N \sim 1/P$, it is found that 
 \begin{equation}
 N \sim  3 \ {\rm x } \ 10^3 \ \Omega_L^{-1} \ 
             \left( {1 \ {\rm AU} \over D_{sep}}\right)
             \left( {M_L \over M_{\odot}}\right)^{1/2}
             \left( {\Delta A \over 0.1 }\right) ,
 \end{equation}
where $D_{sep}$ is the separation of satellites, and $M_L$ is 
the mass of the lenses, and $N$ is the number of objects that 
must be observed before one object would be expected to show 
a parallax lensing effect above $\Delta A$.  
 
Optimistically, there are scenarios involving high densities, 
low lens masses, and high precision measurements where 
detection is virtually assured.  Also, universes involving
high $\Lambda$ may yield small $N$.  Although these universes might 
not be preferred, their existence can be tested with this method.
 Pessimistically, millions of simultaneous measurements of
cosmological objects to moderate precision must be made with expensive spacecraft to detect a single parallax lensing effect created by the known star field, and/or a comparable density of MACHOs.  

Let's reverse the above question and ask how accurate does 
relative simultaneous photometry have to be to see lensing parallax
with a single observation?  Here it is assumed that $u$ is large so that 
amplification is low, very near unity. At very large $u$, it is 
clear from equation 4 that, in magnitude, 
 \begin{equation}
 | {dA \over du} | \sim  { 8 \over u^5 }.   
 \end{equation}
To find a probable measurement, set $P \sim 1$ in equation 7, 
substitute $\Phi = u^2/2$, solve for $u$, and substitute the result in the above equation, so that  
 \begin{equation}
  \Delta A \sim  8 \ 3^{5/2} \ \Omega_L^{5/2}
     \left( { D_{sep} \over E_O }  \right) ,
 \end{equation} 
which becomes, when $z_S = 1$ and stated in more intuitive relative quantities,
 \begin{equation}
  \Delta A \sim  10^{-5} \ \Omega_L^{5/2}
        \left( { D_{sep} \over 1 \ {\rm AU} } \right)
        \left( { M_{\odot} \over M_L } \right)^{1/2}  .
 \end{equation}
$\Delta A$ would be about a factor of 10 more likely were the source
at $z_S = 2$, and an additional factor of 3 more likely were the source
at $z_S = 3$.  Still, the above equation confirms that lensing parallax is practically undetectable, for present technology, for paired 
observations of a single source.  Therefore, this method hinges on 
many sources being observed.

How could two separated observers best measure a small 
differential magnitude of a variable source?   One possible route 
is for each observer to simultaneously monitor the same field,
which includes a common set of comparison objects.   Simultaneous 
observations should minimize the effect of target intrinsic 
variability.  Comparison observations should minimize 
systematic errors of independent magnitude estimation.  The 
precise effectiveness of this minimization will determine the 
usefulness of this method.

Which sources are best candidates for lensing parallax?  To 
assess this, the angular size of a source will be estimated from 
its minimum variability time scale, and by assumption of a 
canonical distance.  When the source is angularly smaller than 
the size of the Einstein ring of the lens, large parallax 
amplifications are possible.  The relationship between source 
variability and the Schwarzschild radius of the lens is
therefore given by
 \begin{equation}
 R_S \ge { ( c \Delta t_{var} )^2 D_{OL} \over
                    2 D_{LS} D_{OS} } .
 \end{equation}
Again applying our canonical values, a more intuitive equation is
 \begin{equation}
 M_L \ge  \left( { \Delta t_{var} \over 1 \ {\rm sec} } \right)^2
           \ 10^{-12.5} \ M_{\odot} .
 \end{equation}

Sources with variability satisfying equation (15) 
make good candidates for gravitational 
parallax observations.  The relative angular size of these 
objects relative to the angular Einstein-ring size of the 
lens makes it possible for two observers to see 
measurably different gravitational lens magnifications.  
Were the object increasing in size, differential 
magnification would cease when the when equations (14) 
and (15) were no longer valid, allowing the observers 
to then estimate the mass of the lens from the size 
and/or variability of the source.

For GRBs in the gamma-ray band, the 
minimum time scale of variability is unknown (Nemiroff et al.,
1998), but significant fluctuation certainly exist in many GRBs above 
$t_{var} \sim 0.01$ sec.  Therefore, parallax 
observations of GRBs could resolve, theoretically, masses above 
$10^{-16} M_{\odot}$ (Nemiroff \& Gould, 1995).  X-ray 
emission is also seen from GRBs over the time scale of hours, 
indicating that lens masses above $10^{-5} M_{\odot}$ could 
be resolved. Variability over the time scales of a day, a  month, 
and a year then could theoretically resolve, through lensing parallax,
lens masses 
greater than about 0.01 $M_{\odot}$, 10 $M_{\odot}$, and 1000 
$M_{\odot}$ respectively. Note that some QSOs and erupting supernovae
are also seen to undergo variability on the time scale of a 
day or less, and so might make good sources for lensing parallax 
observations with lenses of sub-solar masses.

Could the transverse speed of the lens affect its parallax 
measurability?  Phrased differently, the above analysis implicitly 
assumed that the lens, observer, and source were stationary
with respect to each other, but significant relative lens motion
during the observation could confound the 
detection of this effect.  When the lens moves on order 
$D_{sep}$, an amplitude change, for one observer, 
on order of that discussed above would occur.  
Projecting all observer, lens, and source motion
onto transverse speed of the lens, $v_L$, a canonical 
duration of this effect can be estimated as 
 \begin{equation}
    t_{dur} \sim 
            \left( { D_{sep} \over v_L } \right) \sim 
                    3.5 \ {\rm days}  \ 
            \left( D_{sep} \over 1 \ {\rm AU} \right)
            \left( 500 \ {\rm km/sec} \over v_L \right) .
 \end{equation}
A single observer would expect lens motion to create a 
$\Delta A$ on the order of that given by equation (13) over this 
time-scale.  A measurable differential amplification between 
observers should persist significantly longer, however.  

\section{Discussion and Conclusions }
 
Gravitational lens induced amplification patterns  
continually cross our solar system.  For a sufficiently compact 
source, stellar and MACHO 
lenses create a literal sea of amplifications on 
which we are adrift.  Usually the lensing waters are smooth, so 
that we do not perceive this sea.  Even when an 
amplification wave comes, it is hard for us to tell it from  
intrinsic source variability on which it is 
superimposed.  But {\it two} sailors adrift on this amplification 
sea should both measure the same intrinsic source variability, 
and so should be able to interpret differences in perceived
brightness more clearly in terms of differential gravitational
lensing. 

Stated differently, this method attempts to subtract unwanted attributes  
from wanted attributes. If an observer determines that 
a source has a certain brightness, the observer does not know 
how much of this 
brightness is intrinsic, and how much is lens amplified.  To 
subtract intrinsic brightness, separated observers are employed.  

If two separated observers each determine a different brightness
for the same source, the observers do not know how much of this brightness
difference was caused by intrinsic variability of the source.
To subtract source variability, simultaneous observations are employed.

If two simultaneous, separated observers each determine
a different brightness for the same source, the observers still do not
know how much of this difference was caused by systematic errors
operating between the observers.  To subtract systematic effects,
relative photometry between the target source and nearby
unlensed comparison sources is employed.  It is possible that what is left 
is relative gravitational lens amplification. 
 
Ideal sources for this method are bright, compact, distant sources.  The 
more compact the source, the greater the variety of compact 
lenses that create a detectable effect.  The brighter the source, 
the better the relative photometry.  The more distant the source,
the greater the chance a lens will fall between it and the observers.
 
Ideal lenses for this method have much less than a solar mass, 
since they create the highest amount of differential amplification 
across our solar system.  MACHOs and/or stars from the known 
star field may also be detectable, depending on the accuracy of 
the relative photometry.

Ideal observers for this method are large, identical, highly 
separated observers.  Large telescopes can track more sources 
and more accurately.  The more similar the two separated 
observers are, the less systematic effects would likely dominate 
their relative photometry.  The more separated the observers are, 
the higher the lens mass this method is sensitive to.
 
One possible set of satellites that might attempt measurements 
of this type are HST and NGST.   Each would need to have  
identical filters to ensure that measured magnitude differences 
were not spectrum based.   One check that measured magnitude 
differences are truly lens based can be done for expanding 
sources.  When an expanding source becomes much larger than 
the Einstein ring of the lens, these relative magnitude differences 
should diminish.  An independently determined expansion rate 
would calibrate the angular size of the Einstein ring and hence
give valuable information on the mass of the lens.

\section{Acknowledgements}
This research was supported, in part, by grants from NASA and 
NSF.

\end{document}